# Reduced Microwave Brightness Temperature in a Sunspot Atmosphere due to Open Magnetic Fields

A. Vrublevskis, B.I. Ryabov, and S.M. White

**Abstract** Motivated by dark coronal lanes in SOHO / EIT 284 Å EUV observations we construct and optimize an atmosphere model of the AR 8535 sunspot by adding a cool and dense component in the volume of plasma along open field lines determined using the Potential Field Source Surface (PFSS) extrapolation. Our model qualitatively reproduces the observed reduced microwave brightness temperature in the northern part of the sunspot in the VLA observations from 13 May 1999 and provides a physical explanation for the coronal dark lanes. We propose application of this method to other sunspots with such observed dark regions in EUV or soft X-rays and with concurrent microwave observations to determine the significance of open field regions. The connection between open fields and the resulting plasma temperature and density change is of relevance for slow solar wind source investigations.

## 1. Introduction

No scientific consensus exists on an effective solar atmosphere model (see Loukitcheva et al. (2017) for an overview). This issue is relevant for the resolution of the well-known coronal heating problem. Active region (AR) sunspot observations in microwave wavelengths can be used for solar atmosphere density and temperature profile investigations. In this paper, for the AR 8535 sunspot we further relate the observed decrease in microwave intensity to open field lines and thereby demonstrate the additional contribution of high angular resolution microwave observations to magnetic field connectivity determination.

Significant contribution to the microwave radiation from large sunspots is from the gyroresonance emission of electrons in the strong (compared to surrounding quiet Sun) sunspot magnetic fields. Two features make this emission of particular use in investigating the active region atmosphere: (1) the emitted radiation is in local thermodynamic equilibrium (LTE) with the plasma hence the brightness temperature derived from the observed electromagnetic radiation flux in the cases of sufficient optical depth represents the actual local electron temperature, and (2) for gyroresonance emission plasma is of significant optical depth only in thin layers where the emission frequency is close to the gyrofrequency or its harmonic. As a result then gyroresonance emission brightness temperature observations represent the plasma temperature at select heights in the atmosphere. A successful "inversion" of the microwave observations could then provide information on the atmosphere and the magnetic fields of sunspots.

---

A. Vrublevskis
Ventspils International Radio Astronomy Centre and Ventspils University of Applied Sciences, Inzenieru iela 101, Ventspils LV-3601, Latvia
e-mail: arturs.vrublevskis@venta.lv

B. I. Ryabov
Ventspils International Radio Astronomy Centre and Ventspils University of Applied Sciences, Inzenieru iela 101, Ventspils LV-3601, Latvia
e-mail: ryabov@latnet.lv

S. M. White
Space Vehicles Directorate, Air Force Research Laboratory, Kirtland AFB, NM, USA
e-mail: stephen.white.24@us.af.mil

Since the realization of the relevance of the gyroresonance emission mechanism to the observed microwave emission from sunspots (Zheleznyakov, 1962; Kakinuma and Swarup 1962), numerous attempts with varying successes have been made in investigating active regions using such observations (for review, see Gary and Keller (eds.), 2004; Lee, 2007). Arguably, the most straightforward approach is to use a model for the sunspot magnetic field (based on either a photospheric magnetic field extrapolation or a dipole model) together with an assumed atmosphere model (usually plane-parallel) to calculate the microwave emission which is then compared to observations. Such approach can be successful for sufficiently simple sunspots yet numerous exceptions exist.

In some cases authors concluded that the field above the sunspot is not a potential field (Alissandrakis, Kundu, and Lantos, 1980; Nindos et al., 1996; Brosius et al., 2002) and a more involved magnetic field model was necessary.

In other cases, it was suggested that a general low temperature plasma volume overlies some parts of the sunspot. In particular, several authors (Alissandrakis and Kundu 1982; Strong, Alissandrakis, and Kundu, 1984; White, Kundu, and Gopalswamy, 1991; Zlotnik, Kundu, and White, 1996; Zlotnik, White, and Kundu, 1998; Bezrukov et al., 2011; Bezrukov, Ryabov, and Shibasaki, 2012) invoked low temperature plasma above the center of the sunspot and at the relevant gyroresonance levels to explain the observed reduction in emission more pronounced than expected due to the small angle between the magnetic field and the line of sight there. Brosius and White (2004) additionally directly observed enhanced transition region EUV emission ("sunspot plume") at the location of reduced microwave emission in coordinated EUV observations with SOHO/CDS.

First Gary and Hurford (1994) and most recently Tun, Gary and Georgoulis (2011) analyzed spatially resolved microwave spectra of active regions without relying on any magnetic field models and discovered locations with spectra with positive slopes - higher brightness temperatures at higher frequencies - that, again, imply lower temperature plasma higher in the solar corona than the underlying high temperature plasma. An important distinction is that in these cases the cool plasma is in loops high in the corona and not necessarily at the relevant gyroresonance levels.

Finally, Vourlidas, Bastian, and Aschwanden (1997) explained microwave observations using a model that encompassed both concepts – lower temperature plasma at certain gyroresonance layers and constrained to particular loops.

For completeness, it should be mentioned that observations in other parts of electromagnetic spectrum – optical and UV spectral lines – are also used to construct sunspot atmosphere models. The "inversion" is more complicated since the radiative transfer then is a non-LTE process and emission depends on the abundance and ionization equilibrium of particular atoms. Nevertheless, numerous models have been constructed and we refer to Loukitcheva et al. (2017) for an overview of the models and their correspondence to radio observations.

Despite the aforementioned advances, research on sunspot atmosphere modelling using microwave observations continues. Goals include better agreement between models and observations as well as justification or elimination of model assumptions. Thus, more recently, Stupishin et al. (2018) iteratively varied a plane-parallel atmosphere profile to use with non-linear force-free reconstructed magnetic field to model RATAN-600 1-D sunspot observations. Also for modelling RATAN-600 observations, Alissandrakis et al. (2019) in turn employed an inversion of the differential emission measure from SDO/AIA observations to obtain the atmosphere profile to be used together with potential field extrapolations.

As for a physics-based justification of the empirical models, Lee et al. (1998) set different temperature profiles for plasma along different magnetic field lines depending on the current along the given field line. More recently, Mok et al. (2016) modelled volumetric plasma heating in an active region as dependent on magnetic field strength, length of field lines, and plasma density. However, there authors simulated and compared to observations radiation in the EUV rather than radio range. Nita et al. (2018) have developed an IDL-based software package `GX_Simulator` in SolarSoft for calculating the expected microwave, X-ray, and EUV emission of model ARs.

In this paper we focus on AR 8535 observed on 13 May 1999 using the VLA radio telescopes. This particular sunspot exhibits features in microwave observations that cannot be reproduced with a simple - plane-parallel - atmosphere model. In particular, the microwave emission is reduced in the northern part of the sunspot and there is an inversion in brightness temperature with respect to frequency in several regions which are brighter in 8 GHz band than in the 5 GHz band. As already described, this is suggestive of cool plasma overlaying hotter lower layers and such an interpretation for this spot was suggested and investigated by Brosius and White (2004) who associated the depression with a sunspot plume detected in coordinated EUV observations with SOHO/CDS. This particular plume was further described in Brosius and Landi (2005). It should be noted that the above authors refer to this active region as AR 8539.

Ryabov and Shibasaki (2016) investigated this same sunspot in microwave, EUV, and X-ray ranges and argued that the observed reduced brightness in microwaves, dark lanes in EUV and X-ray images, regions of open field lines, and regions of outflows all overlap and are related. Their interpretation was that hot coronal plasma had been evacuated along the open field lines leaving a volume of cooler plasma that is seen as a depression in the microwave observations. Their work extends previous work by Ryabov et al. (2015) where authors provided similar arguments to explain reduced microwave brightness at 17 GHz frequency in the peripheral areas of five other select sunspots.

Our goal in this paper is to model the sunspot atmosphere containing such low temperature plasma along open field lines to obtain an agreement with microwave observations. Plasma outflow along open field lines then would have significant observational consequences and would be a physical mechanism to include in justifying empirical models.

While information on the atmosphere and magnetic fields of active regions is of interest in itself, our hypothesis is of additional significance since it implies that active regions contain open field regions that can act as sources of slow solar wind with observational signatures in microwaves. The rich surrounding coronal magnetic field structure of the AR 8535 sunspot is investigated in detail in Ryabov and Vrublevskis (2020) with evidence presented from solar wind modelling and measurements in favor of outflows. The reader is referred to Abbo et al. (2016) for a comprehensive review of observations and current theories regarding sources of slow solar wind.

This paper is organized as follows. In Section 2 we present the observations followed by a description of our modelling of the sunspot atmosphere to match microwave data in Section 3. We then draw conclusions and discuss the results in Section 4.

## 2.     Observations

NOAA active region 8535 was observed on 13 May 1999 with multiple instruments. The active region included a positive polarity sunspot that is the object of present investigation. During the observations the sunspot was located to the northwest of the disk center. Overview data from the MDI instrument (Scherrer et al., 1995) at 24:00 UT and the EIT instrument (Delaboudinière et al., 1995) at 19:06 UT (both aboard the SOHO satellite) are presented in Figure 1.

In the left panel we show the white light image from the MDI instrument. We use it to mark the sunspot umbra and penumbra. The obtained contours are then overlaid for reference in panel (b) and in the following figures. Contours from the MDI white light data at 19:06 UT (not shown) were used in panel (c).

The middle panel displays the measured photospheric longitudinal magnetic field. The maximum value is 2760 G, which is sufficient for relevant gyroresonance harmonic layers to be present in the sunspot atmosphere.

The right panel contains the image of EUV 284 Å observations with the EIT. Of particular interest is the dark region emanating northward from the sunspot suggestive of absence of hot coronal plasma.

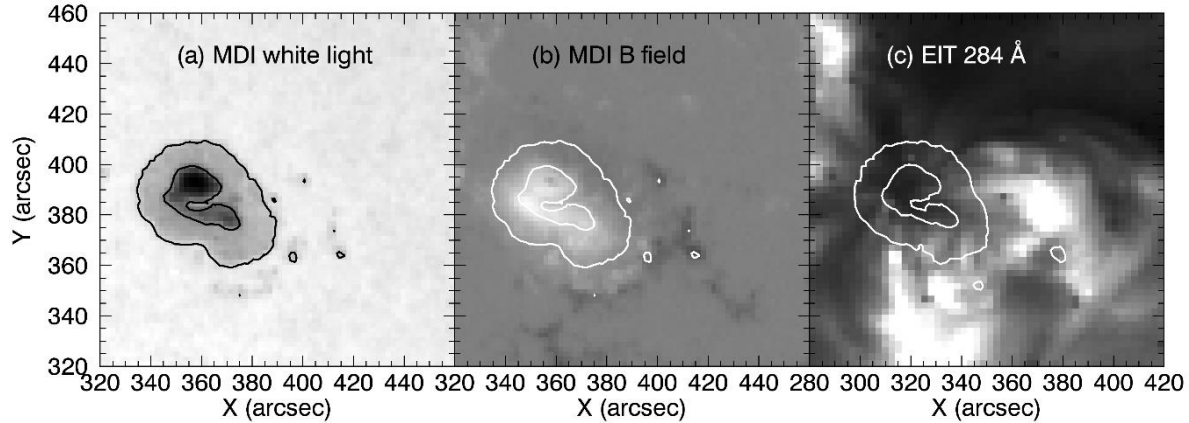

**Figure 1** AR 8535 observed on 13 May 1999. (**a**) MDI white light image and (**b**) MDI photospheric longitudinal magnetic field measurements taken at 24:00 UT. (**c**) EUV 284 Å image from EIT at 19:06 UT. Contours of the sunspot umbra and penumbra are overlaid in panels (b) and (c).

Importantly this sunspot was also observed using the Very Large Array (VLA) radio telescopes as an interferometer between 20:00 and 24:00 UT. Observations at three frequencies – 4.535, 8.065, and 14.665 GHz – and in both the right circular and the left circular polarizations are presented in Figure 2 where the microwave brightness temperature is plotted. Since the sunspot is of positive polarity then the right and left circular polarizations predominantly correspond to correspondingly the extraordinary and ordinary modes of microwave emission. With the VLA telescopes in the "D" configuration the spatial resolution (beam width) for the three frequencies is estimated at 15'', 10'', and 5'' respectively. Note the different color bar scale for the 15 GHz data. The R mode images in panels (a-c) are clipped and the actual measured brightness temperatures reach 3.2, 3.3, and 1.6 MK correspondingly (see Section 3.1.). For better alignment with model data, which are based on MDI magnetic field measurements from 24:00 UT, the VLA data have been shifted 33.4'' to the west and 2.0'' to the south.

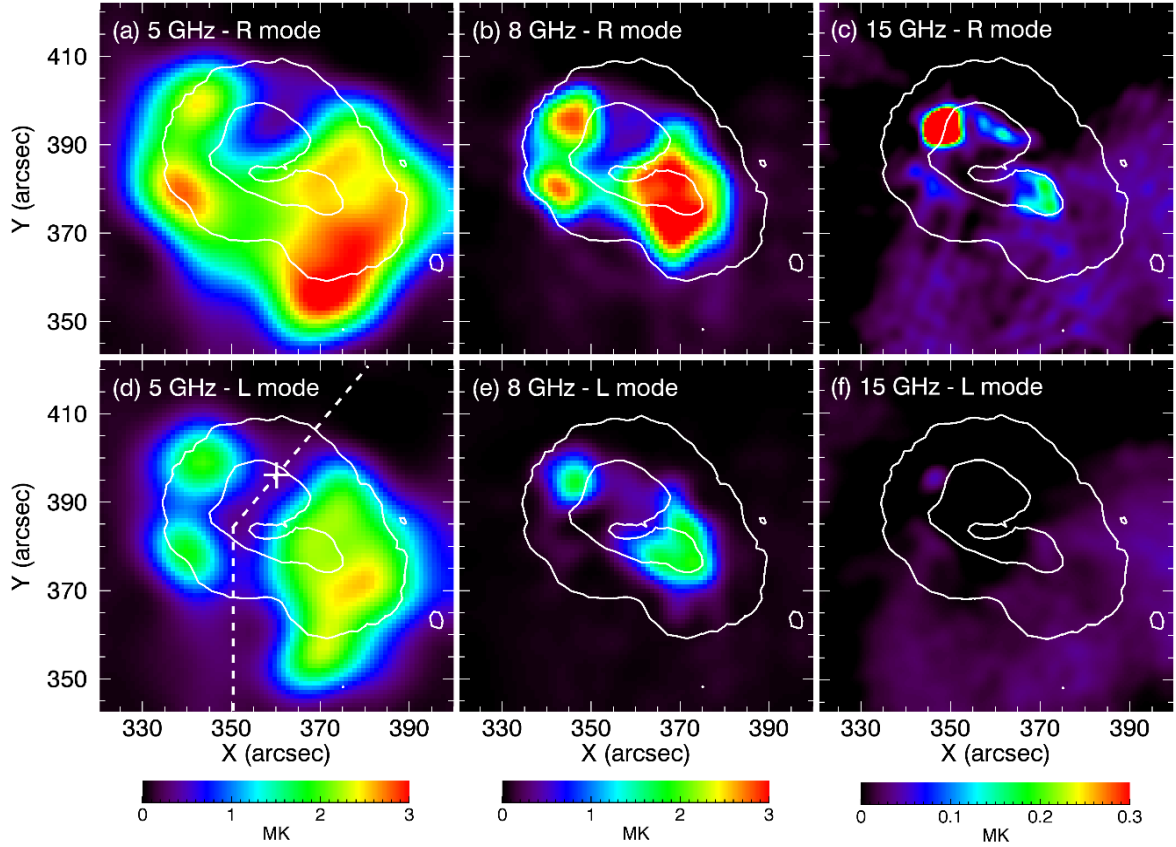

**Figure 2** Brightness temperatures of AR 8535 measured on 13 May 1999 using VLA at three frequencies and in both the right (R) and left (L) circular polarizations. The dashed line in (d) shows the cross cut used in quantitative comparisons with model results. The cross in (d) marks the location that is further analyzed in the text where R mode brightness temperature larger in 8 GHz than in 5 GHz. Contours of the sunspot umbra and penumbra are overlaid in panels (b) and (c).

## 3. Modelling of the Microwave Observations

### 3.1. MODEL SETUP

From Figure 2 the following peculiar observational features that inform the proposed model can be noted:
  I. The sunspot displays an irregular shape in the microwave data. Most clearly in the 8 GHz R mode measurements three bright areas can be distinguished that can be traced to the 5 GHz data as well. The sunspot is particularly irregular in the 15 GHz R-mode data where a 1.6 MK bright spot exists to the north-east from the sunspot center while the brightness temperature is less than 0.3 MK in the rest of the sunspot;
  II. A pronounced region of reduced brightness temperature can be observed in the northern part of the sunspot around (360'', 400'') in the 5 and 8 GHz R mode data;
  III. Four areas exist where the sunspot brightness temperature in the R mode in 8 GHz exceeds the brightness temperature in 5 GHz (blue in Figure 10a). Three of these areas correspond to the bright areas described as part of feature I above. The fourth area near (360'', 396'') (location marked with a cross in Figure 2d) is within the reduced brightness temperature area that is feature II.

We first focus on the three bright areas from features I and III. The higher brightness temperature at 8 GHz than at 5 GHz is unusual since in a typical sunspot the 5 GHz optically thick gyroresonance layers would lie higher in the atmosphere where there is higher electron temperature than at the corresponding 8 GHz gyroresonance layers. As was already pointed out in the Introduction, one possible explanation as invoked by Gary and Hurford (1994) and more recently Tun, Gary and Georgoulis (2011) is that cool coronal loops overlay these areas. Since the optical depth for free-free emission by a given cool loop plasma volume is inversely proportional to the square of the frequency then 5 GHz emission from lower and hotter layers is absorbed more than the 8 GHz emission. However, it is then a special coincidence that these cool loops from feature III are overlaying exactly the areas from feature I that are of significantly higher brightness temperatures than other areas of the sunspot.

An alternative explanation adopted here is that in this particular case at the locations of the bright areas in feature I the plasma actually is not hotter at the higher 5 [GHz] gyroresonance layers than at the lower 8 GHz gyroresonance layers. We adopt this explanation without explicitly establishing the cause for this temperature inversion. The image sizes of the bright areas are comparable to the beam widths at these frequencies. Thus while it cannot be determined with certainty, the locations of the bright areas in the images at different frequencies are consistent with an interpretation that each bright area represents a flux tube that is observed at different heights in different frequencies. Presumably then due to some process plasma is of increased electron temperature along three flux tubes (feature I) with greater temperature at the lower heights of the 8 GHz gyroresonance layers (feature III). Our goal is to model the presumably unperturbed by this process sunspot atmosphere outside these flux tubes.

Unusual is also feature II. The optical depth of gyroresonance emission has strong dependency on the angle between the magnetic field and the line of sight and approaches zero as this angle is decreased. Thus for a sunspot with symmetric magnetic field a decrease in brightness temperature is expected closer to the disk center, which in our case is to the southeast from the sunspot center. The opposite is observed whereby the decreased brightness temperature is to the northwest of the sunspot center.

Inspired by the EUV observations (Figure 1c), similar soft X-ray observations (not shown), and previous work by Ryabov et al. (2015) and Ryabov and Shibasaki (2016), we pose the hypothesis that open magnetic field lines are present in the northern part of the sunspot. Due to increased plasma transport along these open field lines one could then expect reduced amounts of hot coronal plasma in the corresponding atmosphere volume.

To model such a case we constructed a model atmosphere volume centered on our sunspot at (360'', 382'') and consisting of 41x41x2401 rectangular cuboid voxels with horizontal spacing of 0.002058 $R_\odot$ (equivalent to 1.9638'' or 1432 km) and vertical spacing of 117 km (equivalent to 0.000168 $R_\odot$). In the vertical direction (Z-axis) the volume is along the line of sight. We aligned the X-axis with the heliocentric west direction and the Y-axis with the heliocentric north direction.

## 3.2. MAGNETIC FIELD MODEL

We employ the Potential Field Source Surface (PFSS) extrapolation of the measured surface magnetic field (Altschuler and Newkirk, 1969; Schatten et al., 1969) to specify the magnetic field at each voxel.

PFSS extrapolation assumes that magnetic field is completely radial at a "source surface" at some radius $R_{SS}$ which is a model parameter traditionally set at $R_{SS} = 2.5\ R_\odot$. However, other values of this parameter have been used to better match observations. Lee et al. (2011) obtained better agreement with observations with $R_{SS} \approx 1.9\ R_\odot$ and $R_{SS} \approx 1.8\ R_\odot$ for the minimum periods of cycles 22 and 23 respectively. More recently, Bale et al. (2019) used $R_{SS}$ as low as $R_{SS} = 1.2\ R_\odot$ in order to reproduce some of the magnetic field features observed in situ by the Parker Solar Probe at 36 to 54 solar radii in October-November, 2018. A lower source-surface increases the volume of the atmosphere with open field lines. With that in mind below we focus on PFSS extrapolations and the resulting atmosphere models with $R_{SS} = 1.8$ $R_\odot$, exploring the $R_{SS} = 2.5$ $R_\odot$ case afterwards.

Since the reduced brightness region (feature II) that we claim to be caused by open field lines is of 0.5 heliospheric degree characteristic size then a high-resolution magnetic field reconstruction is necessary to model and trace magnetic field lines in our model atmosphere volume. At the same time, a global model is necessary to accurately identify field lines as open or closed. We achieve these dual goals by following these steps:

1. We start with the precalculated 384 x 192 data point global surface radial magnetic field map from LMSAL (Schrijver and Derosa, 2003) that is based on 6-hour cadence observations at 00:04 UT on 14 May 1999. We interpolate the field to a finer 3874 x 1937 data point grid which is roughly double the spatial resolution of the 96-min. cadence full-disk MDI longitudinal magnetic field measurements.
2. We use such 96-min. cadence longitudinal field measurements from 24:00 UT on 13 May 1999 (see Figure 1b) and calculate the radial magnetic field for 161 x161 data points from the sun disk corresponding to the area between 242'' and 557'' along X-axis and between 264'' and 579'' along Y-axis. This high-resolution radial field is likewise interpolated to the grid from step 1 and is used as a substitution for the lower-resolution radial field data near our sunspot.
3. We perform PFSS extrapolation using appropriately modified routines from the `pfss` package (Schrijver and Derosa, 2003) in SolarSoft. Magnetic field is calculated on a grid in spherical coordinates with uniform $\pi / 1937 = 0.00162$ rad spacing in angular coordinates and the same 117 km (equivalent to 0.000168 $R_\odot$) spacing in the radial coordinate for a volume that encompasses the model atmosphere volume. Crucially, the number of spherical harmonics used is kept large in order to ensure high-resolution magnetic field reconstruction. In particular, we implement the same somewhat arbitrary limit as in the SolarSoft `pfss` package routines that the number of spherical harmonics $\ell$ used for reconstructing the field at a radial coordinate r (expressed in $R_\odot$) is such that $r^\ell < 10^6$ and no larger than 1937.
4. We use trilinear interpolation to obtain magnetic field values for the model atmosphere volume. Here it is assumed that, due to the fine coordinate spacing compared to the curvature radius, for interpolation purposes each voxel in spherical coordinates can be approximated as rectilinear.
5. For the global field calculations, we use the default routines and the grid from the LMSAL precalculated fields only with extrapolations performed using our modified surface radial magnetic field map and with our chosen $R_{SS}$ values.

Since the observed gyroresonance emission is strongly dependent on the angle between the magnetic field and the observer then in explaining the reduced microwave brightness temperature (feature II) one must differentiate between the effects of the proposed cool plasma volume and of any smaller scale field irregularities with field aligned closer to the line of view than expected for a regular sunspot. To address this issue in Figure 3b for the atmosphere model volume we have plotted the magnetic field magnitude at a relevant height of 1400 km with contours (going inwards) for harmonics: 540 G for 5 GHz s=3; 810 G for 5 GHz s=2; 960 G for 8 GHz s=3; 1441 G for 8 GHz s=2. The contours are to a great extent symmetric suggesting that magnetic field irregularities are not present to a sufficient extent for the magnetic field structure alone to account for the decreased brightness temperature in the northern part of the sunspot.

The general existence of open field lines can be demonstrated by tracing representative field lines from the surface in the global magnetic field model as shown in Figure 3a where the green field lines are open while the black are closed. In order to identify - using the global model - voxels in our atmosphere model volume that lie on open magnetic field lines we first choose a reference height in the atmosphere model volume where we would relate the high-resolution atmosphere model volume field connectivity to the global model field connectivity. We choose 90 Mm above the sunspot center. Then for each voxel in the atmosphere model volume we trace a field line starting from that voxel and determine the pixel (if any) at the reference height that the field line intersects. For the global model, we trace field lines from coordinates corresponding to each of the pixels at the reference height and determine if the field line is open or closed. Then all of the voxels that are magnetically connected to the reference height pixels in the atmosphere model volume that are in turn determined to lie on open field lines in the global model are considered to also lie on open field lines.

For magnetic field line tracing purposes we used a model volume extended westward and northward to 81x81x2401 pixels in order to properly identify as open field lines that leave the nominal atmosphere model volume.

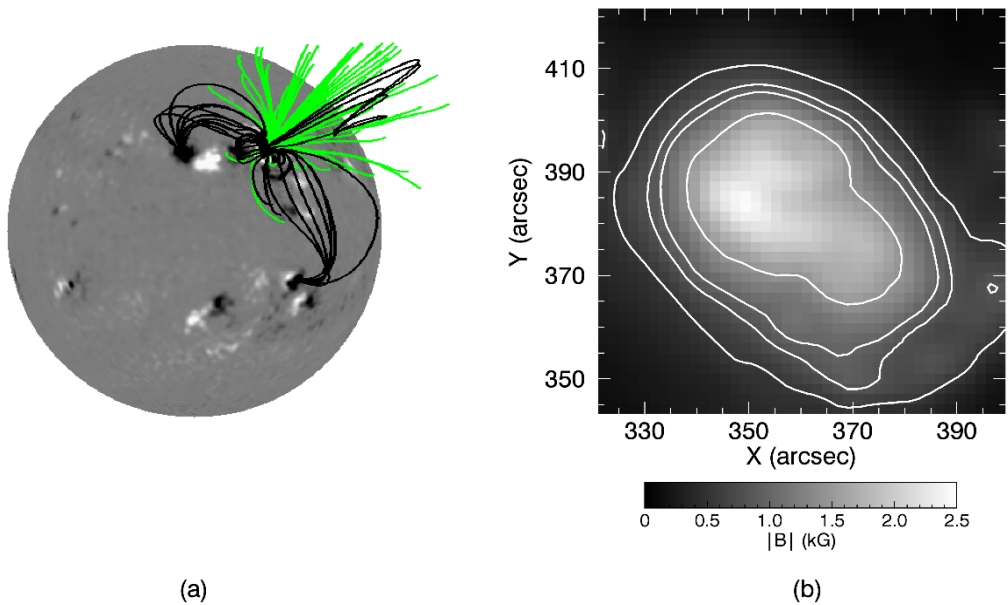

(a)  (b)

**Figure 3** (**a**) Indicative open and closed field lines from select points in AR 8535. (**b**) Magnetic field magnitude at the relevant height of 1400 km with contours for magnetic field magnitudes corresponding to gyroresonance harmonics (going inwards): 5 GHz s=3; 5 GHz s=2; 8 GHz s=3; 8 GHz s=2.

Overall, we divide all of the atmosphere model volume into two sub volumes or components. One – the open field component – corresponds to the voxels lying along open field lines. The other – the bulk component – corresponds to the rest of the volume. One can envision distinguishing additional components in order to model the flux tubes corresponding to the bright areas in the VLA data. The open field component is shown in Figures 4a and 4b from different viewpoints with respect to the atmosphere model volume, the Sun surface, and the sunspot. Note that Z-axis is not to scale in Figure 4a. In Figure 4c borders of the open field component at different atmosphere model volume heights along Z-axis are shown. Near the surface of the Sun at Z = 40 Mm the border of the open field component (solid line in Figure 4c) is in the northern part of the sunspot atmosphere and is remarkably coincident with the region of reduced brightness temperature (feature II) in the microwave observations (see Figures 2a and 2b). The open field lines initially continue northward and the open field component occupies only the northern part of the model volume at Z = 120 Mm (dash-dotted line in Figure 4c). With increasing height the open field lines spread out and eventually a significant portion continue southward across the model volume. At Z = 200 Mm the open field component occupies already half of the model volume cross-section.

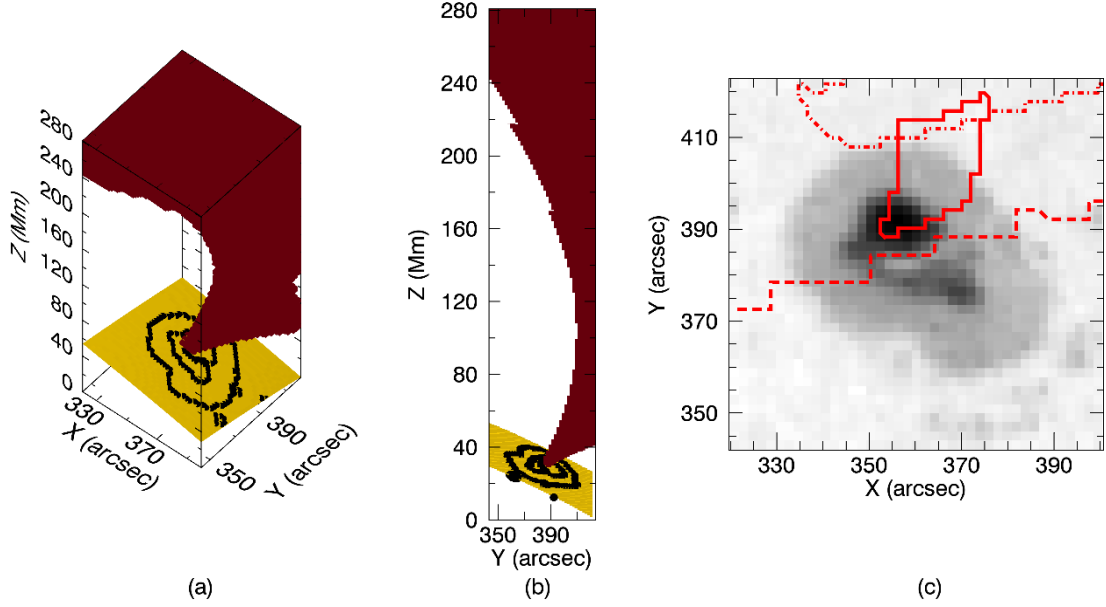

**Figure 4** (**a**)-(**b**) The model atmosphere volume shown from different viewpoints. Dark red - the open field component, yellow - Sun surface, black contours - the sunspot umbra and penumbra. Z-axis is not to scale in (a). (**c**) Borders of the open field sub volume at different atmosphere model volume heights along Z-axis: solid – 40 Mm; dash-dotted – 120 Mm; dashed – 200 Mm. Background: MDI white light image for reference.

### 3.3. ATMOSPHERE MODEL

Having established the two sub volumes, we then set atmosphere parameters – electron temperature and density – as a function of height h above Sun surface for each of the components. We use the model of Alissandrakis, Kundu, and Lantos (1980) that parametrizes the atmosphere above the chosen transition temperature of $T_0$ = 100 000 K using three parameters: (1) height $h_0$ for this transition temperature, (2) pressure at this transition point, and (3) conductive heat flux $F_C$ in the atmosphere above the transition point. For convenience in our case we vary the related quantity - electron density $n_0$ - instead of pressure. Above the transition height the temperature increase is assumed as that for constant conductive heat flux $F_C$ while for density we assume hydrostatic equilibrium and ignore variation of specific gravity with height. We allow the temperature to increase up to a maximum coronal temperature $T_{COR}$ = 8 MK. The exact upper limit is not important since corona becomes optically thin at these heights.

Quantitatively in cgs units we then have (Alissandrakis, Kundu, and Lantos, 1980):

$$T(h) = \left[T_0^{7/2} + \frac{7}{2}\frac{F_C}{A}(h - h_0)\right]^{\frac{2}{7}} \quad (1)$$

and

$$n(h) = n_0 \left(\frac{1}{t(h)}\right) \exp\left[-\frac{282}{F_C}\left(t(h)^{5/2} - 1\right)\right], \quad (2)$$

where A = 1.1 × $10^{-6}$ erg $cm^{-1}$ $s^{-1}$ $K^{-7/2}$ and t(h) = T(h) / $T_0$ .

At heights below $h_0$ we assume constant temperature of 5000 K and density of $10^{11}$ $cm^{-3}$.

With magnetic field and plasma temperature and density defined in the model volume we calculate the expected brightness temperature at the three observational frequencies. We model two kinds of emission mechanisms – thermal gyroresonance and free-free emission (i.e. thermal bremsstrahlung). We adapted code from the FORWARD package (Gibson et al., 2016) in SolarSoft. The opacity κ at frequency f due to free-free emission is calculated approximately according to the simplified formula (Dulk, 1985; Gelfreikh, 2004):

$$\kappa = 0.2 \frac{n^2}{T^{3/2}(f \pm f_B |\cos\theta|)^2}, \qquad (3)$$

where $f_B$ is the gyrofrequency at the given voxel location and θ is the angle between the magnetic field direction and the line of sight. The plus sign is for the ordinary mode, the minus – the extraordinary. From opacity the optical depth dτ due to a layer of geometrical depth ds can be calculated according to:

$$d\tau = \kappa\, ds \qquad (4)$$

For thermal gyroresonance, instead the optical depth due to a gyroresonance layer is directly calculated for voxels where these layers are crossed. In the FORWARD package this is done using the approximations from Robinson and Melrose (1984). Effectively then within our model the width of all gyroresonance layers is 117 km. We should also note that we modified the FORWARD routines and for the magnetic field scale length estimate used the derivative along the line of sight of the magnetic field magnitude B rather than of the component of the field along the line of sight $B_z$: $L_B = B \times |dB/dz|^{-1}$ instead of $L_B = B_z \times |dB_z/dz|^{-1}$.

The total optical depth at the given frequency for the given voxel is the sum of the optical depths due to both mechanisms. Knowing the optical depth, the radiative transfer equation is integrated along each of the 41x41 vertical columns within the atmosphere model volume. Passing through each voxel some of the incident radiation with brightness temperature $T_B$ is absorbed while some with the voxel electron temperature $T_e$ is emitted depending on the optical thickness of the layer dτ and according to the formula:

$$T_B' = T_B e^{-d\tau} + T_e(1 - e^{-d\tau}), \qquad (5)$$

where $T_B'$ is the brightness temperature of the radiation exiting the given voxel.

After calculating the brightness temperatures for each individual column we then apply Gaussian smoothing to account for beam width for a comparison with the observed data to be made.

For a given atmosphere model the fit to observations is evaluated by calculating the $\chi^2_{pol,f}$ for each polarization and frequency as well as the combined $\chi^2$ according to formulas:

$$\chi^2_{pol,f} = \frac{1}{\left[T^{B,obs}_{pol,f}\right]_{max}} \sum_i \left(T^{B,obs}_{pol,f}(i) - T^{B,model}_{pol,f}(i)\right)^2 \qquad (6)$$

and

$$\chi^2 = \sum_{pol,f} \chi^2_{pol,f}, \qquad (7)$$

where "pol" stands for either the R or the L polarization and "f" for one of the three observational frequencies. The comparison and the summation is done for image points indexed with "i". "$T^{B,\,obs}$" refers to the brightness temperature image from observations while "$T^{B,\,model}$"

– to the corresponding image from the atmosphere model. For the given frequency and polarization "[ ]$_{max}$" designates the maximum brightness temperature observed among all of the image points indexed with "i" and used in the comparison.

Since the observed images include bright areas (feature I) that we do not model, then for the model fit evaluation instead of using the complete image we choose just a cross cut starting south of the sunspot, directed first northward towards the center of the sunspot, and then continuing to the north-west across the region of reduced brightness temperature (feature II). The cross cut is illustrated with a dashed line in Figure 2d. The southern part of the cross cut would correspond to the sunspot atmosphere largely unperturbed by the open field line region while the northern part would be representative of it. At 5 GHz we chose a total of 46 points spaced 1.9638'' apart along the cross cut. To account for the higher spatial resolution in the 8 GHz and 15 GHz observations, the number of points are 69 and 138 with 1.309'' and 0.655'' spacing respectively.

In the model of Alissandrakis, Kundu, and Lantos (1980) for our atmosphere model volume which consists of two components a total of six parameters would be necessary to specify the temperature and density throughout the volume. However, we assume that the height $h_0$ of the transition temperature is the same for both components set by horizontally uniform processes below this height rather than affected by the plasma outflow along the open field lines in the corona. This is a simplifying assumption since it ignores the Wilson depression. Within our model for the optimized parameters afterwards we checked the impact of varying the height $h_0$ for the open field component and found the effect insignificant.

Qualitatively, the open field component must be modelled with low temperature in order for the brightness temperatures to be lower than in the bulk component. This necessitates smaller conductive heat flux values so that the temperature increase with height is more gradual and remains low in the heights effectively sampled by the gyroresonance mechanism. For the density of the open field component, both low and high values can potentially produce the desired effect and must be explored – lower values would lead to lower brightness temperatures through absence of plasma while higher values can be beneficial if higher brightness temperature emission from lower atmosphere layers is to be absorbed at higher layers.

We initially calculated a total of 52 000 models. For the bulk component we explored 10 different $n_0$ values (in cm$^{-3}$ log $n_0$ between 7.4 and 11), 8 $F_C$ values (in erg cm$^{-2}$ s$^{-1}$ log $F_C$ between 5.4 and 8.2), and 10 $h_0$ values between 800 km and 2600 km. For the open field component: 13 different $n_0$ values (in cm$^{-3}$ log $n_0$ between 7.4 and 12.2) and 5 $F_C$ values (in erg cm$^{-2}$ s$^{-1}$ log $F_C$ between 0.0 and 1.6). For these parameters the best agreement was obtained for log $F_C$ = 0.0 for the open field component which is effectively an isothermal plasma. In such case the exact value of the chosen transition temperature $T_0$ gains additional significance and must be explored.

We thus calculated additional 864 models. For the bulk component we varied parameters near the minimum already found while for the open field component we fixed log $F_C$ = 0.0 and varied $T_0$ between 8 different values 60 000 to 200 000 K. The explored and fitted parameter values are listed in Table 1.

In Figure 5a the model atmosphere temperatures and densities with height for the bulk and open field components using the fitted parameters values are plotted while in Figures 5b and 5c respectively the temperature and density along the cross cut used in the fitting are shown. The cut is not along a straight line (see Figure 2d) and the location where the direction of the cut changes is shown with a vertical dashed line at Y = 384''. Z-axis is not to scale. The contours in black represent (with decreasing height) gyroresonance layer locations for harmonics: 5 GHz s=3; 5 GHz s=2; 8 GHz s=3; 8 GHz s=2.

| Component | Parameter | Value range | Fitted value |
|---|---|---|---|
| Bulk | | | |
| | log $n_0$ (cm$^{-3}$) | 8.2; 8.6; 9.0 | 8.6 |
| | log $F_C$ (erg cm$^{-2}$ s$^{-1}$) | 6.2; 6.6; 7.0 | 6.6 |
| | $h_0$ (km) | 1200; 1400; 1600 | 1400 |
| | $T_0$ (K) | 100 000 | 100 000 (fixed) |
| Open field | | | |
| | log $n_0$ (cm$^{-3}$) | 9.0; 9.4; 9.8; 10.2 | 9.4 |
| | log $F_C$ (erg cm$^{-2}$ s$^{-1}$) | 0.0 | 0.0 (fixed) |
| | $h_0$ (km) | 1400 | 1400 (fixed) |
| | $T_0$ (K) | 60 000; 80 000; …; 200 000 | 140 000 |

**Table 1** The parameter ranges and the fitted with respect to the observations values for the atmosphere model bulk and open field components.

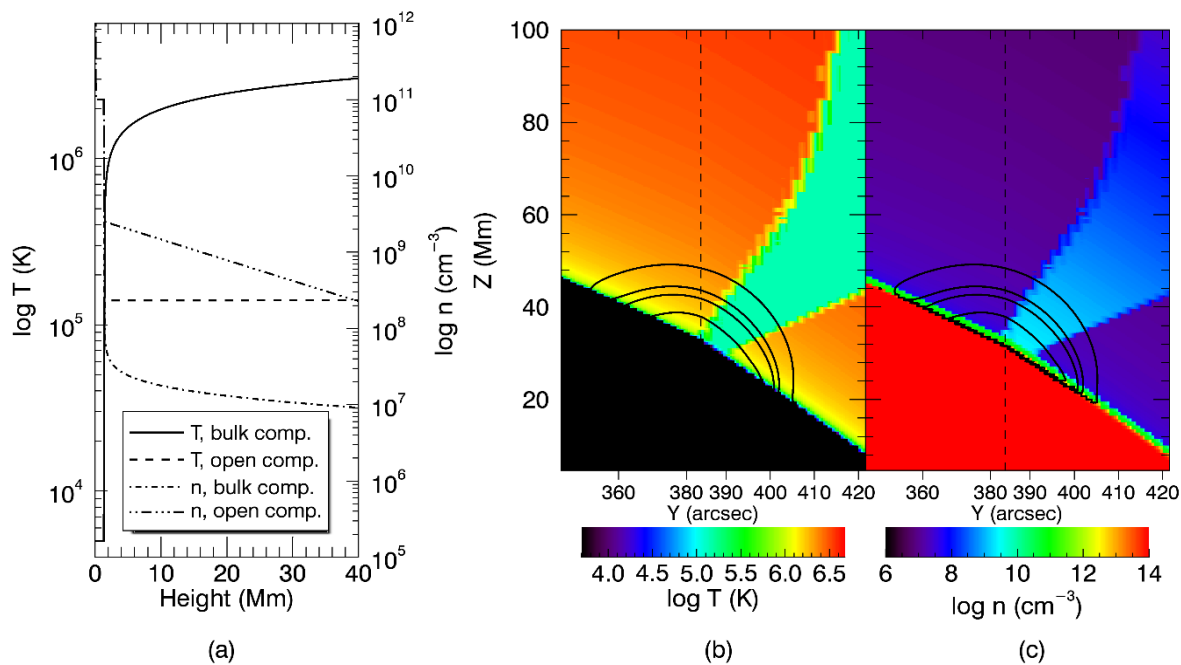

**Figure 5 (a)** Temperature and density with height for the bulk and open field components. **(b)** and **(c)** Temperature and density respectively along the cross cut used in the fitting. The contours in black represent (with decreasing height) gyroresonance layer locations for harmonics: 5 GHz s=3; 5 GHz s=2; 8 GHz s=3; 8 GHz s=2. Z-axis is not to scale.

### 3.4. MODEL RESULTS

Below we present the atmosphere model results using the fitted with respect to the observations values from Table 1. Figures 6 and 7 show the model brightness temperature images at the frequencies and polarizations of the VLA observations. In Figure 6 Gaussian smoothing to account for VLA beam widths has been applied and represents the data used in the quantitative comparison with the observations as described above. Figure 7 is the same data before smoothing useful for analysis below.

Figure 8 presents in more detail the observed and model brightness temperatures along the cross cut shown in Figure 2d and used in the fitting. As in Figure 5, the vertical dashed line marks the location where the direction of the cut changes. The solid line corresponds to the observed temperatures, the dashed - to the atmosphere model. For comparison, also shown with a dotted line are the brightness temperatures in an atmosphere model where all of the volume is parametrized as a bulk component only leading to a completely plane-parallel single-component model. The model brightness temperatures images for this single-component model are shown in Figure 9.

We note the following regarding the model results in comparison to observations:
1. As can be seen in Figures 6a,b and 8a,b the model reproduces a decrease in brightness temperature in R mode at 5 and 8 GHz in the northern part of the sunspot. The location as judged from the images before smoothing (Figure 7a,b) is in agreement with the observations. However, in the smoothed images the area of reduced brightness temperature is less pronounced than observed (Figure 2a,b).
2. At the same image area in R mode at 15 GHz before smoothing the model (Figure 7c) leads to a relatively small increase of 7000 K compared to the single-component model (not shown) in the maximum brightness temperature. The increase is even less with smoothing applied. In observations (Figures 2c, 8c), a pronounced peak of nearly $T^{B,\,obs}_{R,\,15\,GHz} = 100\,000$ K is observed in the northern part of the sunspot (Figure 8c) compared to $T^{B,\,model}_{R,\,15\,GHz} = 33\,000$ K in the model after smoothing.
3. The single-component model does not adequately reproduce the observations – there is no brightness temperature decrease in the northern part of the sunspot. Instead, as theoretically expected, brightness temperature is lower in the southeast part of the sunspot towards the Sun disk center (Figure 9d) due to the smaller angle between the magnetic field and the line of sight there. This furthers supports the claim already made based on the magnetic field magnitude contours at 1400 km above the Sun surface in Figure 3b that any magnetic field irregularities are not present to a sufficient extent for the magnetic field structure alone to account for the decreased brightness temperature in the northern part of the sunspot within plane-parallel atmosphere models.
4. The discrepancies between the model and observed brightness temperatures in the southern part of the sunspot (see Figure 8c,d) suggest that even with the use of a cross cut the impact of the bright areas on perturbing the brightness temperature is not completely avoided.

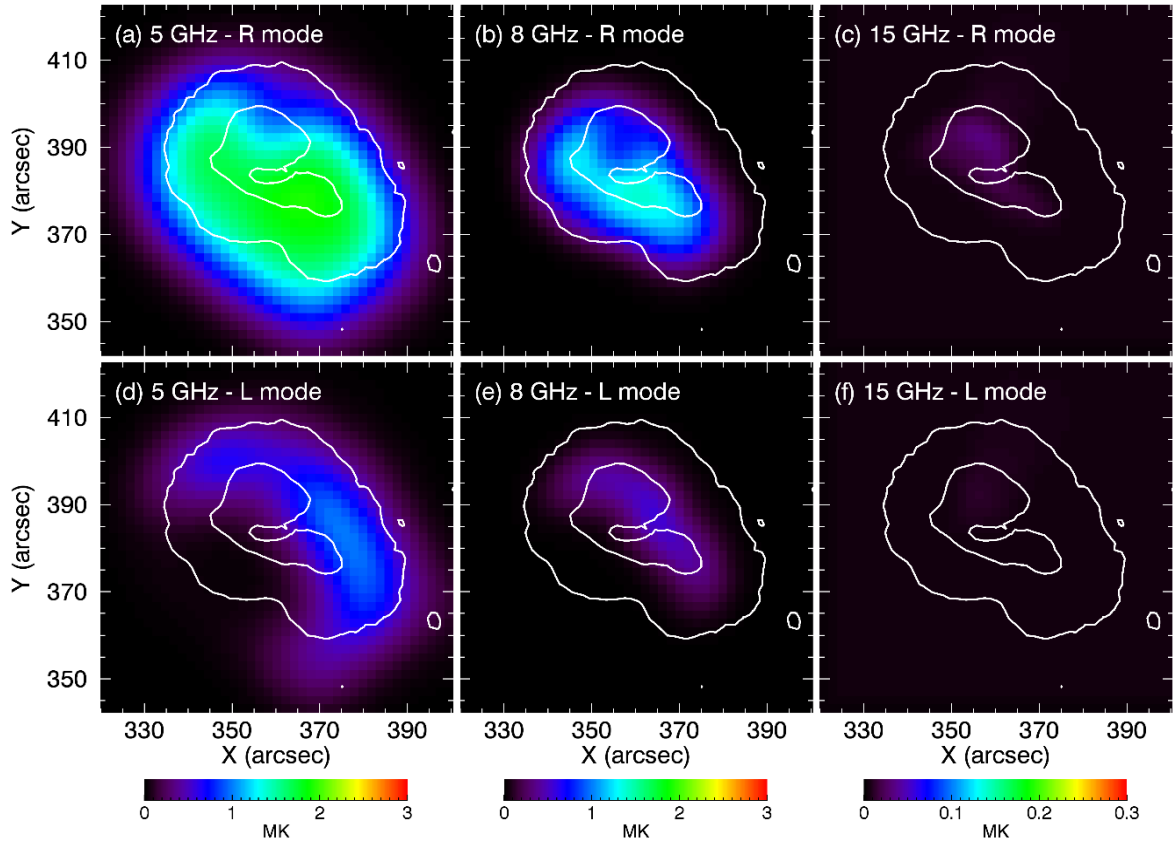

**Figure 6** Model brightness temperature images at the three VLA observation frequencies and in both the right (R) and left (L) circular polarizations. Overlaid are contours of the sunspot umbra and penumbra.

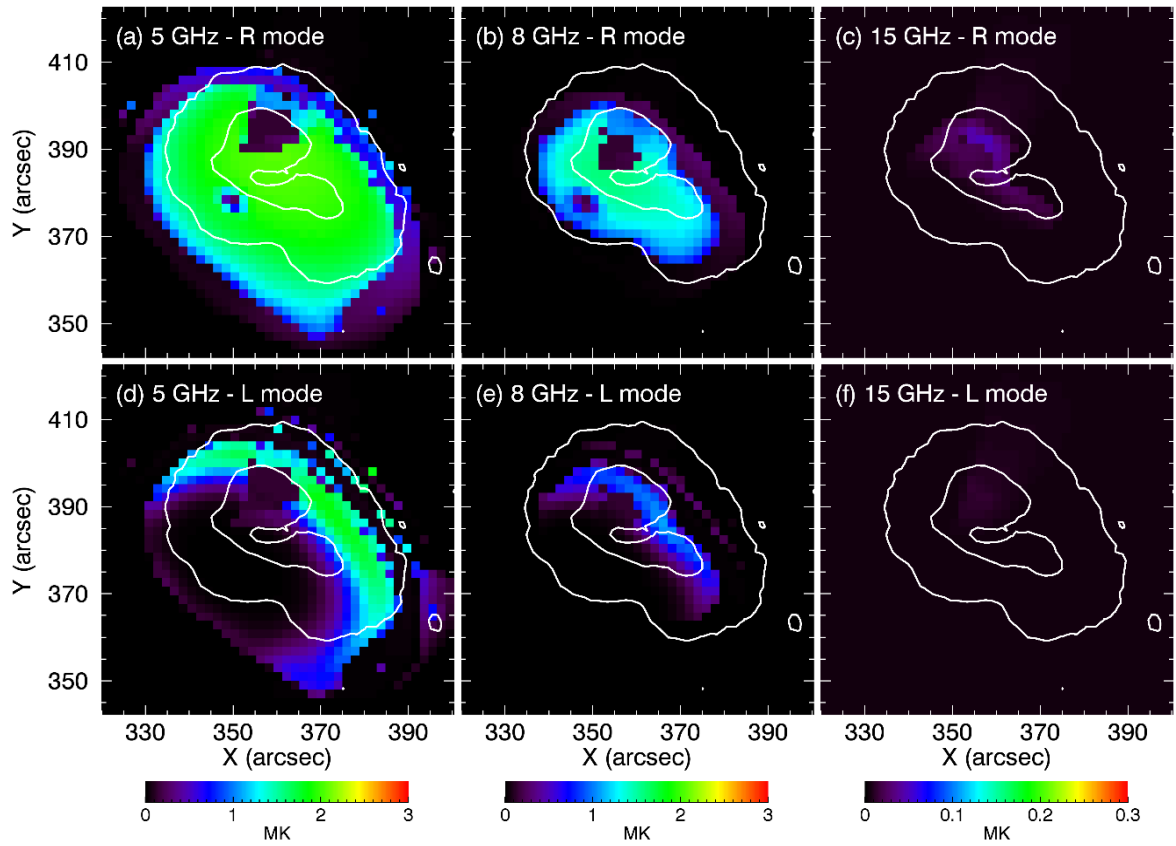

**Figure 7** Model brightness temperature images at the three VLA observation frequencies and in both the right (R) and left (L) circular polarizations before the application of the Gaussian smoothing to account for VLA beam widths. Overlaid are contours of the sunspot umbra and penumbra.

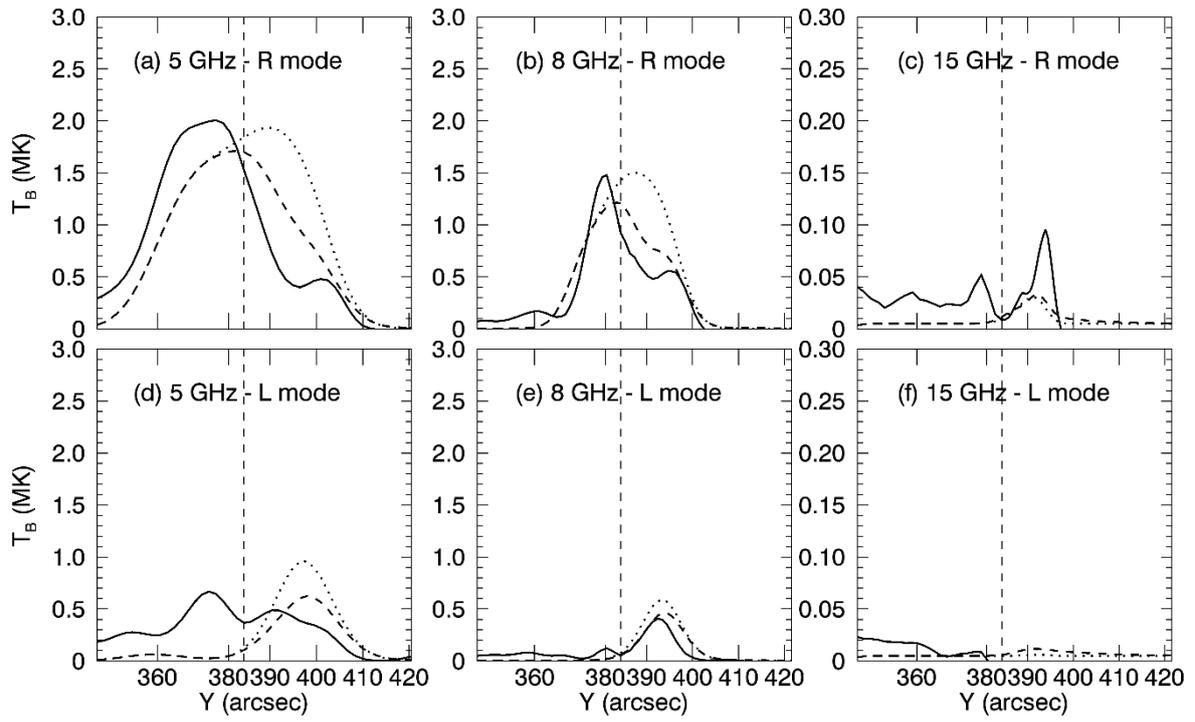

**Figure 8** The observed (solid) and model (dashed) brightness temperatures at the three VLA observation frequencies and in both the right (R) and left (L) circular polarizations along the cross cut used in the fitting. Also shown (dotted) the brightness temperatures for a single-component plane-parallel model.

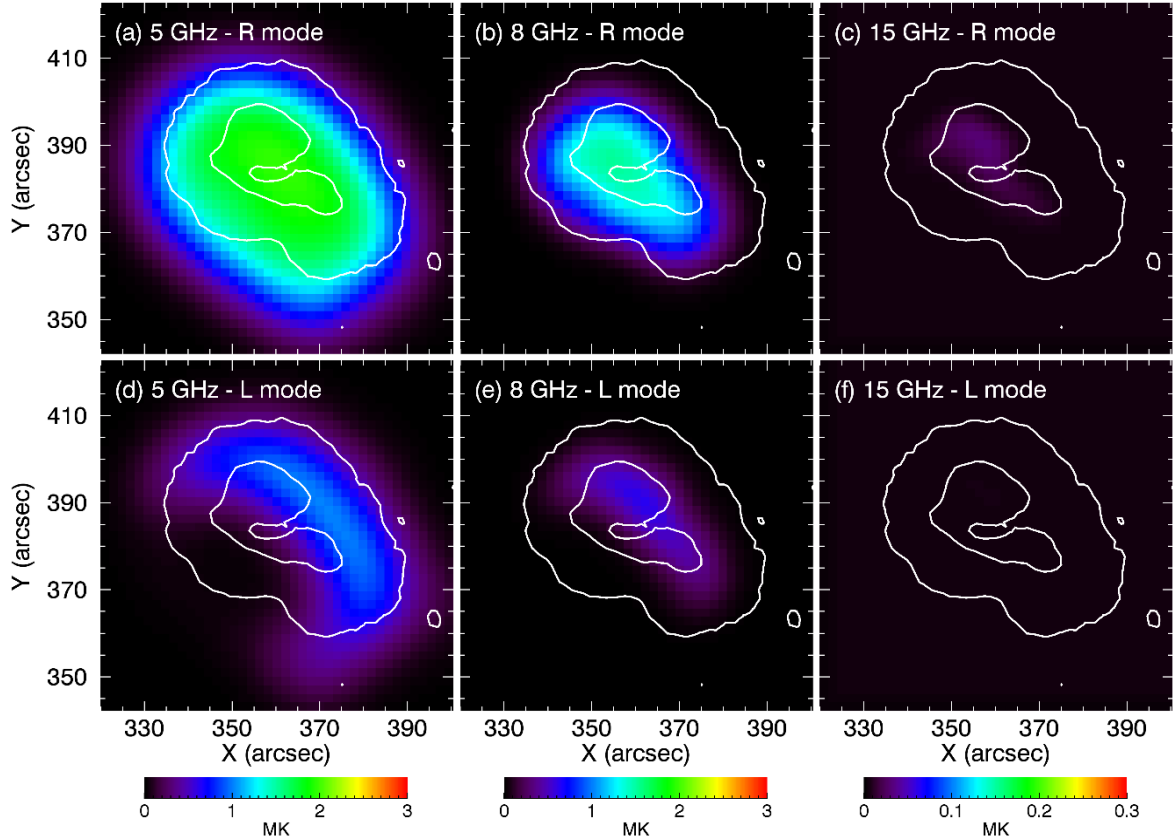

**Figure 9** Model brightness temperature images at the three VLA observation frequencies and in both the right (R) and left (L) circular polarizations for a single-component plane-parallel model. Overlaid are contours of the sunspot umbra and penumbra.

In the observational data in Section 2 we noted feature III – an area near (360'', 396'') (location marked with a cross in Figure 2d) within the broader reduced brightness temperature area in the northern part of the sunspot where the sunspot brightness temperature in the R mode in 8 GHz exceeds the brightness temperature in 5 GHz. In Figure 10 we show the difference $T^B_{R,\,5\,GHz} - T^B_{R,\,8\,GHz}$ between the R mode brightness temperature in 8 GHz and in 5 GHz for data from: (a) the VLA observations; (b) our model before the application of the Gaussian smoothing in accordance with VLA beam widths; and (c) our model after smoothing. Qualitatively – before smoothing – our model reproduces the brightness temperature inversion at this location. However, as with the overall reduced brightness temperature area in the northern part of the sunspot, the application of the Gaussian smoothing to account for the beam width diminishes the effect.

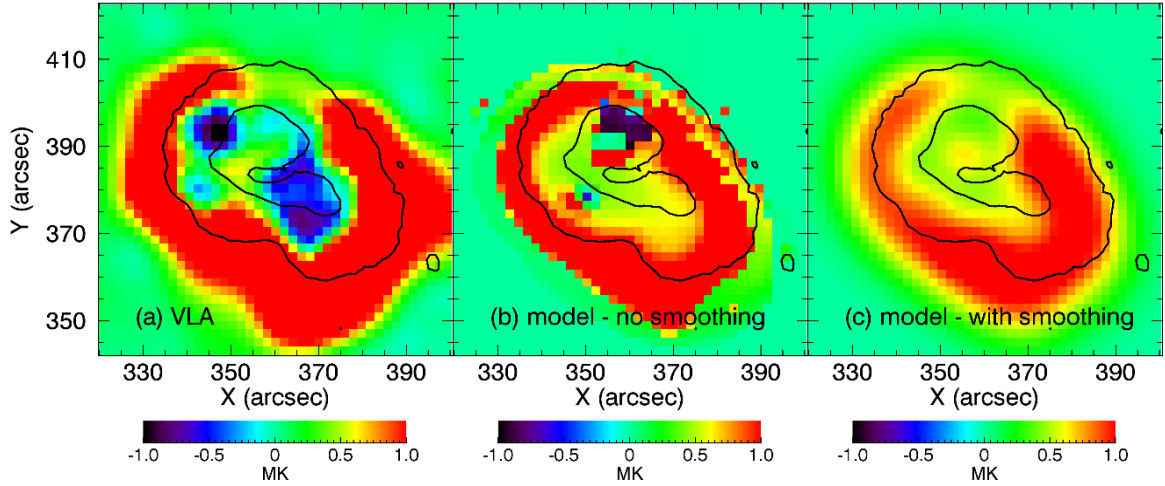

**Figure 10** Images of the difference between the R mode brightness temperature in 8 GHz and in 5 GHz for data from: **(a)** the VLA observations; **(b)** the model atmosphere before Gaussian smoothing in accordance with VLA beam widths; and **(c)** the model atmosphere after smoothing. Overlaid are contours of the sunspot umbra and penumbra.

For this same location (marked with a cross in Figure 2d) of brightness temperature inversion with frequency within the area of reduced brightness temperature, in Figure 11 we investigate in more detail the formation with height of the model brightness temperatures which then allows for establishing the physical reasoning for the model results and the obtained fitted atmosphere models. Shown in the panels with respect to height along Z-axis above the surface of the Sun are (from top): electron density; temperature; integrated from the surface according to Eq. 5 brightness temperatures in the VLA observation frequencies and polarizations; the free-free (dashed) and thermal gyroresonance (solid) opacities in R mode at the three VLA observation frequencies. The vertical dashed lines mark locations of gyroresonance layers as listed on the top.

Focusing on the R mode as in Figure 10, it can be seen from Figure 11a,b that in our model at this location the open field component lies at heights between 8 and 27 Mm above the surface of the Sun. The relevant $s = 2$, 3, and 4 gyroresonance layers at 5 GHz sample the low temperature open field component leading to the corresponding brightness temperature of $T^{B,\,model}_{R,\,5\,GHz} = 140\,000$ K at this sunspot location before image smoothing. At 8 GHz the $s = 2$ harmonic is at the height of 3 Mm and is optically thick within the hot bulk component. As a result, the brightness temperature reaches 1 MK. The optical thickness both of the higher gyroresonance layers lying within the open field component as well as due to the free-free emission in the dense plasma is insufficient to significantly decrease the brightness temperature with height and it remains at $T^{B,\,model}_{R,\,8\,GHz} = 900\,000$ K before smoothing. At 15 GHz the model brightness temperature is set by the free-free emission within the dense open field component leading to a steady increase with height to $T^{B,\,model}_{R,\,15\,GHz} = 12\,000$ K.

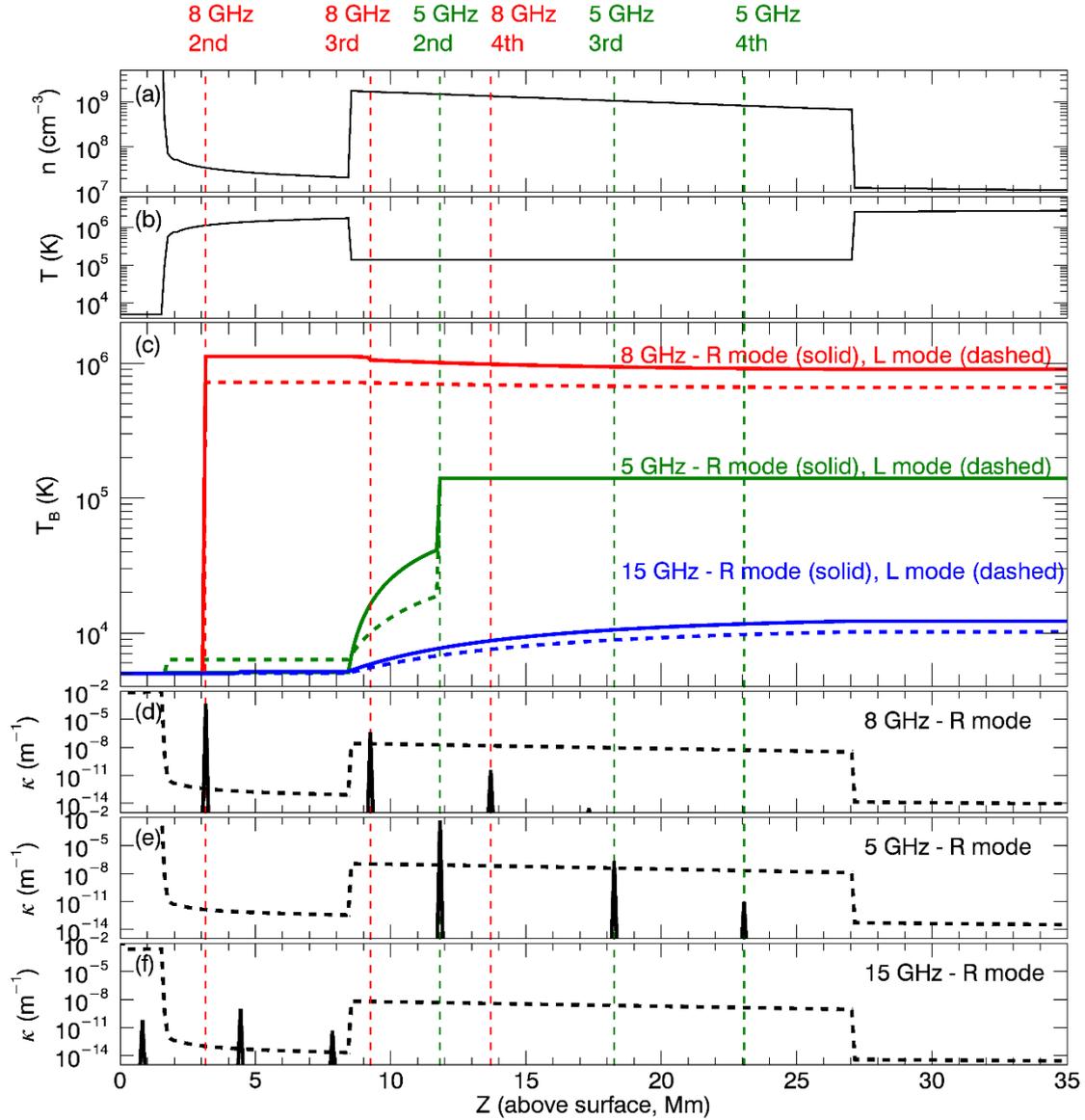

**Figure 11** Model **(a)** electron density, **(b)** electron temperature, **(c)** integrated from the surface brightness temperatures in the VLA observation frequencies and polarizations, and the free-free (dashed) and thermal gyroresonance (solid) opacities in R mode at the three VLA observation frequencies **(d)-(e)** with height along Z-axis above the surface of the Sun near (360'', 396'') at the location of inverted with frequency brightness temperature. The vertical dashed lines mark locations of gyroresonance layers as listed on the top.

The physical explanation for the existence of an optimal fit with respect to the electron density and temperature of the open field component is as follows:
1. The model brightness temperature decrease in the northern part of the sunspot predominantly due to optically thick gyroresonance emission from gyroresonance layers within the open field component will be more pronounced the smaller is the electron temperature there.
2. Within the model of Alissandrakis, Kundu, and Lantos (1980) – Eqs. 1 and 2 above - an isothermal temperature profile leads to a density profile that is exponentially decreasing with height (see Figure 5a) where the exponent is inversely proportional to temperature. Thus, a lower temperature leads to a more rapidly decreasing density.
3. Further north of the previously investigated location, now within the area of reduced brightness temperature in penumbra – near (360'', 400'') - the open field component starts sufficiently high in the atmosphere (see Figures 4a,b and 5b) that the 5 GHz s = 2 harmonic layer is located lower in height and within the hot bulk component. This leads to large brightness temperature that is then reduced due to free-free emission within the dense open field component. Since opacity increases with density then the density must be sufficiently large to significantly reduce the brightness temperature and for the reduced brightness area thus extend north beyond the sunspot umbra and include penumbra.

4. Since at 15 GHz the brightness temperature is determined by the free-free emission from the dense open field component, then larger density leads to an increase in model brightness temperature at 15 GHz in the northern part of the sunspot. At even larger densities such bright area however extends northward even beyond the sunspot penumbra in disagreement with observations.

The above listed four counteracting effects lead to an optimal fit electron temperature and density for the open field component within our atmosphere model.

As described, an increase in the open field density leads to an increase in the brightness temperature at 15 GHz. Thus, within our model it is possible to qualitatively reproduce the observed (Figures 2c, 8c) brightness temperature increase in the northern part of the sunspot at 15 GHz. Within the fitting framework used, however, the optimal fit taking into account all frequencies and polarizations is at a lower density with a peak in brightness temperature smaller than observed: $T^{B, obs}_{R, 15 GHz}$ = 100 000 K is observed compared to $T^{B, model}_{R, 15 GHz}$ = 33 000 K in the model after smoothing.

As already was noted in the Introduction, Brosius and White (2004) and Brosius and Landi (2005) observed a plume at the location of the reduced microwave emission for this sunspot in coordinated EUV observations with SOHO/CDS. For the plume they report temperatures between $1.6 \times 10^5$ and $5.0 \times 10^5$ K. In our model the best fit is obtained with a slightly smaller temperature of $1.4 \times 10^5$ K. As for the density then the previously reported value for the logarithm of the electron density is 9.4±0.2. In our model the density of the open field component exponentially decreases from log $n_0$ = 9.4 with the characteristic values of log $n_0$ ~ 9 at the relevant heights.

The above analysis used a magnetic field model based on a PFSS extrapolation with $R_{SS}$ = 1.8 $R_\odot$. The source-surface was purposefully chosen lower than the standard $R_{SS}$ = 2.5 $R_\odot$ since it leads to an increase in the open field line atmosphere volume and the features of reduced brightness temperature in model images due to the open field component already are almost unresolved after smoothing. Below in Figure 12 we illustrate the results of using a magnetic field model based on a PFSS extrapolation with $R_{SS}$ = 2.5 $R_\odot$ and the same fitted atmosphere parameters from Table 1. Figure 12a is the same as Figure 4c only for an $R_{SS}$ = 2.5 $R_\odot$ extrapolation. Open fields are still present and at the same location within the sunspot only here at the lower heights (solid line) the volume is narrower than with $R_{SS}$ = 1.8 $R_\odot$. The open field lines again initially continue northward, spread out with increasing height and eventually a significant portion continue southward across the model volume (not shown). However, this takes place at larger heights as is expected with an increased source-surface height

Figures 12b and 12c show the model brightness temperature images before and after smoothing respectively in R mode at 5 GHz with $R_{SS}$ = 2.5 $R_\odot$ magnetic field extrapolation and with the same fitted atmosphere parameter values as for $R_{SS}$ = 1.8 $R_\odot$. Qualitatively the atmosphere model still produces reduced brightness temperature area in the northern part of the sunspot at the correct location of the observational feature II. However, the area is even narrower than in the $R_{SS}$ = 1.8 $R_\odot$ model before smoothing (compare Figures 12b and 7a) and the effect is barely noticeable in the image after smoothing.

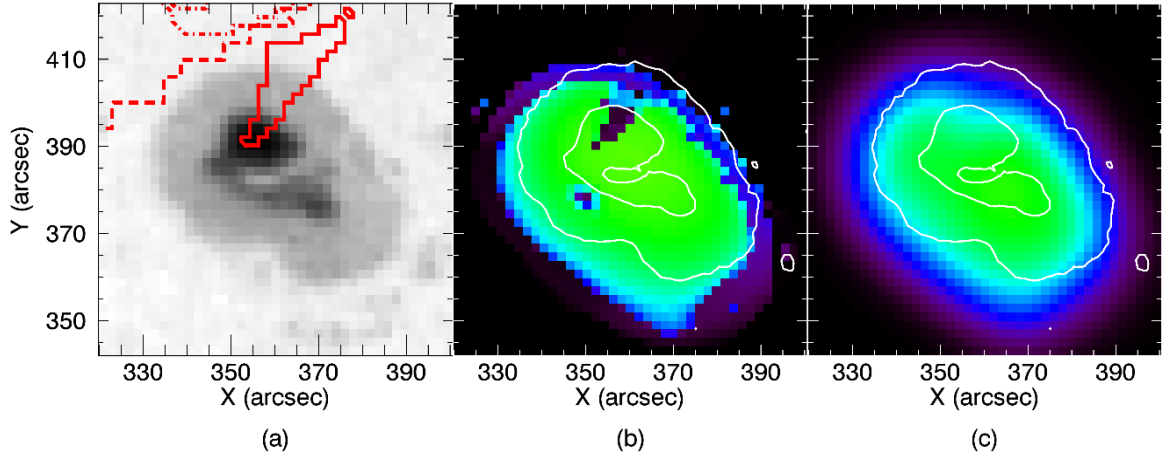

**Figure 12** For the magnetic field extrapolation with $R_{SS}$ = 2.5 $R_\odot$: **(a)** borders of the open field sub volume at different atmosphere model volume heights along Z-axis: solid – 40 Mm; dash-dotted – 120 Mm; dashed – 200 Mm. Background: MDI white light image for reference; **(b)** and **(c)** model brightness temperature images in R mode at 5 GHz before and after smoothing to account for the VLA beam width. Overlaid are contours of the sunspot umbra and penumbra.

## 4.  Discussion and Conclusions

We have constructed a model volume for the atmosphere above the sunspot of AR 8535. We performed high-resolution PFSS reconstruction to determine the magnetic field for the volume while also identifying an open magnetic field line sub volume. We use the model of Alissandrakis, Kundu, and Lantos (1980) to parametrize the atmospheres of the two sub volumes or components – the bulk component and the open field component. For the latter to achieve better agreement with observations we further explored the subcase of isothermal plasma and varied the temperature parameter. The bulk component determines the overall size and temperature of the sunspot in microwaves. Varying the density and temperature of the open field component produces opposing effects in different frequencies and sunspot areas thus allowing for optimization. We emphasize the following results and conclusions:

1. In our high resolution magnetic field modelling we identify an open field line volume spatially coincident with the location of reduced brightness temperature in the VLA 5 and 8 GHz R mode images (compare Figures 4c and 12a with Figures 2a,b).
2. Our model that includes a dense low-temperature isothermal component in the open field line sub volume qualitatively accurately reproduces the brightness temperature decrease in the northern part of the sunspot in the VLA 5 and 8 GHz R mode images.
3. The reduced brightness temperature feature in the model images becomes significantly less distinct and is less pronounced in comparison to the observations after Gaussian smoothing to account for VLA beam width has been applied to the images. This suggests that the atmosphere sub volume with dense low temperature plasma is broader than modeled and was the motivation for performing the PFSS magnetic field extrapolation with the source-surface at $R_{SS}$ = 1.8 $R_\odot$. As stated in Section 3.2., while such a choice is not unprecedented, the required broader low temperature plasma volume beyond that predicted by the standard $R_{SS}$ = 2.5 $R_\odot$ modelling may be instead due to cold, dense or intermediate between the two components plasma along closed loops neighboring the open field volume. As shown in Figure 12, modelling with $R_{SS} = 2.5\ R_\odot$ still reproduces the open field volume and reduced brightness temperature in the northern part of the sunspot though not as extensive as observed.
4. The sunspot microwave images contain three areas of increased brightness temperature that also correspond to areas of brightness temperature inversion with frequency. Due to this coincidence and likely magnetic linkage between gyroresonance layers for the different frequencies corresponding to these bright areas we postulate that physically these represent flux tubes with higher electron temperature at higher atmosphere layers. This is different from the cases of Gary and Hurford (1994) and most recently Tun, Gary and Georgoulis (2011) where brightness temperature inversion with frequency was due to cool overlaying loops. While here we do not further elaborate on the nature or source of these flux tubes such an investigation is warranted including a relation, possibly through magnetic reconnection, to the neighboring open fields that we establish.
5. The atmosphere model bulk component alone leads to brightness temperature images (Figure 9) that do not reproduce the brightness temperature decrease in the northern part of the sunspot. As is also suggested by the featureless magnetic field magnitude contours in Figure 3b, then it is unlikely that the brightness temperature decrease is due to magnetic field irregularities alone and is not reproducible with a single-component plane-parallel atmosphere model.
6. We conclude that the at least for this particular sunspot of AR 8535 the low temperature plasma that had already been observed and described generally as a sunspot plume (Brosius and White, 2004) is also consistent with the alternative interpretation of low temperature plasma lying along open field lines.

We did not attempt to model the areas of increased brightness temperature – feature I from Section 3.1. Hence, for the model fit evaluation we used cross cuts instead of the complete images. However, Figures 8c,d suggest that the impact of the bright areas on perturbing the brightness temperature has not been completely avoided. More complete accounting for the bright areas may allow for more accurate model fitting.

The presented model is empirical and limited. While within the employed atmosphere model both bulk and open field components are in hydrostatic equilibrium, no equilibrium considerations between the two components are investigated. Likewise, no flows are modeled even though such may be of importance with open field lines.

Our present investigation was limited to a single sunspot but we propose repeating the analysis for other sunspots with radio observations in several frequencies that have either been determined to contain a low temperature plasma volume (see Introduction) or that display dark lanes in EUV and X-ray images.

The connection between open fields and the resulting plasma temperature and density decrease is of relevance for slow solar wind source research and suggests further investigation of plasma flows in the atmosphere of this sunspot and with respect to our presented atmosphere structure.

Overall, we demonstrate that radio images of the sunspot of AR 8535 provide a wealth of information regarding its atmosphere which in our interpretation contains flux tubes with higher temperature higher in the atmosphere and open field lines with dense low temperature plasma.

**Acknowledgements** Magnetic field and EUV data provided by the SOHO MDI and EIT consortia. SOHO is a project of international cooperation between ESA and NASA. Karl G. Jansky Very Large Array (VLA) is a component of the National Radio Astronomy Observatory, which is a facility of the National Science Foundation operated under cooperative agreement by Associated Universities, Inc. A.V. is supported by the ERDF Postdoctoral research aid project No. 1.1.1.2/16/1/001 research application No. 1.1.1.2/VIAA/1/16/079 «Understanding Solar Magnetic Atmosphere» (USMA). A.V. and B.I.R. are thankful to the Ventspils City Council for support.

**Figure captions**

**Figure 1** AR 8535 observed on 13 May 1999. (**a**) MDI white light image and (**b**) MDI photospheric longitudinal magnetic field measurements taken at 24:00 UT. (**c**) EUV 284 Å image from EIT at 19:06 UT. Contours of the sunspot umbra and penumbra are overlaid in panels (b) and (c).

**Figure 2** Brightness temperatures of AR 8535 measured on 13 May 1999 using VLA at three frequencies and in both the right (R) and left (L) circular polarizations. The dashed line in (d) shows the cross cut used in quantitative comparisons with model results. The cross in (d) marks the location that is further analyzed in the text where R mode brightness temperature larger in 8 GHz than in 5 GHz. Contours of the sunspot umbra and penumbra are overlaid in panels (b) and (c).

**Figure 3** (**a**) Indicative open and closed field lines from select points in AR 8535. (**b**) Magnetic field magnitude at the relevant height of 1400 km with contours for magnetic field magnitudes corresponding to gyroresonance harmonics (going inwards): 5 GHz s=3; 5 GHz s=2; 8 GHz s=3; 8 GHz s=2.

**Figure 4** (**a**)-(**b**) The model atmosphere volume shown from different viewpoints. Dark red - the open field component, yellow - Sun surface, black contours - the sunspot umbra and penumbra. Z-axis is not to scale in (a). (**c**) Borders of the open field sub volume at different atmosphere model volume heights along Z-axis: solid – 40 Mm; dash-dotted – 120 Mm; dashed – 200 Mm. Background: MDI white light image for reference.

**Figure 5** (**a**) Temperature and density with height for the bulk and open field components. (**b**) and (**c**) Temperature and density respectively along the cross cut used in the fitting. The contours in black represent (with decreasing height) gyroresonance layer locations for harmonics: 5 GHz s=3; 5 GHz s=2; 8 GHz s=3; 8 GHz s=2. Z-axis is not to scale.

**Figure 6** Model brightness temperature images at the three VLA observation frequencies and in both the right (R) and left (L) circular polarizations. Overlaid are contours of the sunspot umbra and penumbra.

**Figure 7** Model brightness temperature images at the three VLA observation frequencies and in both the right (R) and left (L) circular polarizations before the application of the Gaussian smoothing to account for VLA beam widths. Overlaid are contours of the sunspot umbra and penumbra.

**Figure 8** The observed (solid) and model (dashed) brightness temperatures at the three VLA observation frequencies and in both the right (R) and left (L) circular polarizations along the cross cut used in the fitting. Also shown (dotted) the brightness temperatures for a single-component plane-parallel model.

**Figure 9** Model brightness temperature images at the three VLA observation frequencies and in both the right (R) and left (L) circular polarizations for a single-component plane-parallel model. Overlaid are contours of the sunspot umbra and penumbra.

**Figure 10** Images of the difference between the R mode brightness temperature in 8 GHz and in 5 GHz for data from: (**a**) the VLA observations; (**b**) the model atmosphere before Gaussian smoothing in accordance with VLA beam widths; and (**c**) the model atmosphere after smoothing. Overlaid are contours of the sunspot umbra and penumbra.

**Figure 11** Model (**a**) electron density, (**b**) electron temperature, (**c**) integrated from the surface brightness temperatures in the VLA observation frequencies and polarizations, and the free-free (dashed) and thermal gyroresonance (solid) opacities in R mode at the three VLA observation frequencies (**d**)-(**e**) with height along Z-axis above the surface of the Sun near (360'', 396'') at the location of inverted with frequency brightness temperature. The vertical dashed lines mark locations of gyroresonance layers as listed on the top.

**Figure 12** For the magnetic field extrapolation with $R_{SS} = 2.5\ R_\odot$: **(a)** borders of the open field sub volume at different atmosphere model volume heights along Z-axis: solid – 40 Mm; dash-dotted – 120 Mm; dashed – 200 Mm. Background: MDI white light image for reference; **(b)** and **(c)** model brightness temperature images in R mode at 5 GHz before and after smoothing to account for the VLA beam width. Overlaid are contours of the sunspot umbra and penumbra.